\documentclass[a4paper]{jpconf}

\usepackage{caption}
\usepackage{subcaption}
\usepackage{graphicx}
\usepackage{url}
\usepackage[colorlinks=true,linkcolor=black,urlcolor=black,citecolor=black]{hyperref}

\begin{document}

\title{Experiment Software and Projects on the Web with VISPA}

\author{M Erdmann, B Fischer, R Fischer, E Geiser, C Glaser, G M\"uller, \newline M Rieger, M Urban, R F von Cube and C Welling}

\address{III. Physics Institute A, RWTH Aachen University, Germany}

\ead{rieger@physik.rwth-aachen.de, rfischer@physik.rwth-aachen.de}

\begin{abstract}
The Visual Physics Analysis (VISPA) project defines a toolbox for accessing software via the web. It is based on latest web technologies and provides a powerful extension mechanism that enables to interface a wide range of applications. Beyond basic applications such as a code editor, a file browser, or a terminal, it meets the demands of sophisticated experiment-specific use cases that focus on physics data analyses and typically require a high degree of interactivity. As an example, we developed a data inspector that is capable of browsing interactively through event content of several data formats, e.g., „MiniAOD“ which is utilized by the CMS collaboration. The VISPA extension mechanism can also be used to embed external web-based applications that benefit from dynamic allocation of user-defined computing resources via SSH. For example, by wrapping the „JSROOT“ project, ROOT files located on any remote machine can be inspected directly through a VISPA server instance. We introduced domains that combine groups of users and role-based permissions. Thereby, tailored projects are enabled, e.g. for teaching where access to student's homework is restricted to a team of tutors, or for experiment-specific data that may only be accessible for members of the collaboration. We present the extension mechanism including corresponding applications and give an outlook onto the new permission system.
\end{abstract}

\section{Introduction}

Scientific workflows can strongly benefit from the advantages of web applications. Especially location-independent access via web browsers and centralized software installation on dedicated servers are convincing arguments. As an example, \textit{IPython Notebooks} \cite{ipython} become more and more popular in the physics community.

However, specialized experiment-specific software requires an entirely different level of infrastructure in order to profit from web-based approaches. The VISPA system is capable of coping with the demands of complex experiment environments in the web \cite{vispa}.

The article is structured as follows.
Section \ref{sec:software} describes the architecture of the VISPA software as well as its key concepts.
Subsequent sections \ref{sec:pxl} and \ref{sec:jsroot} present developed extensions that demonstrate its suitability as an environment for experiment-specific software.
In section \ref{sec:um} the user management and permission system is explained which introduces configurability of resources and access to them.

\section{The VISPA Software}\label{sec:software}

Following a server-client approach, the VISPA software consists of two parts.
The server is written in Python and is implemented using the \textit{CherryPy} framework \cite{cherrypy}.
The browser-based client is written in JavaScript with graphical components defined in HTML5 \cite{html5} and CSS3 \cite{css3}.
Communication between the two relies on \textit{Hypertext Transfer Protocol} (HTTP) requests and \textit{WebSockets}.
User-related information, i.e. login data, permissions and preferences, are stored in a relational database, which is accessed via \textit{object-relational mapping} (ORM) using the \textit{SQLAlchemy} toolkit \cite{sqlalchemy}.
The internal user management system enables administrators to group users into logical entities, such as \textit{Projects} and \textit{Groups}, and handles allocation of permissions.

Two key concepts are the ability to connect to user-defined resources, called \textit{workspaces}, and a mechanism that allows for extending functionalities.
Both of them are described in detail in the following.

\subsection{Workspace Architecture}

\begin{figure}[h!tbp]
\begin{center}
\vspace{-3mm}
\includegraphics[width=0.7\textwidth]{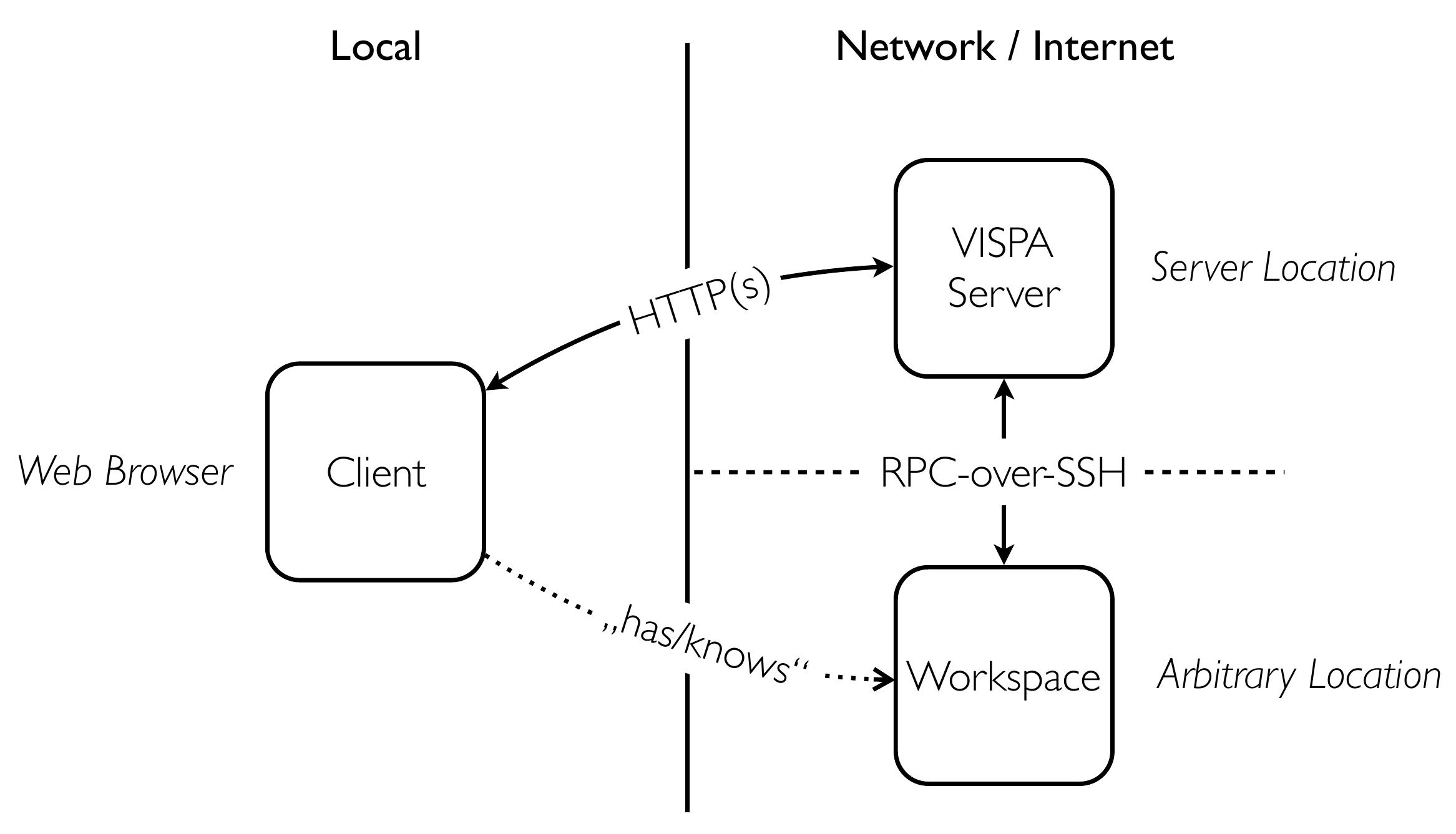}
\caption{Schematic overview of the workspace architecture \cite{acat13}.}
\label{fig:arch1}
\vspace{-5mm}
\end{center}
\end{figure}

The purpose of the VISPA server is to provide a working environment for multiple users.
Thus, in order to preserve server responsiveness, requests for large amounts of data and computing power should be treated separately.
Workspaces constitute the means to including dynamic and scalable resources into the server-client architecture.
A schematic overview is shown in figure \ref{fig:arch1}.

A workspace is a worker node that runs a Python interpreter and is accessible through a network via SSH.
These requirements are met by numerous different types of machines: desktop computers, computing clusters, and even tablets and smartphones.
Resource utilization and file access are granted as the server relies on the user login to interact with workspaces via \textit{remote procedure calls} (RPC).
The connection between server and clients as well as between server and workspaces is encrypted via SSL and SSH, respectively.

Using the graphical interface of the client, users can connect to predefined workspaces and define and connect to custom workspaces.

\subsection{Extension Mechanism}

Often experiment-specific software is specially tailored for its particular use case.
Any system or environment that aims to cover several use cases simultaneously should exhibit appropriate flexibility.
Within VISPA, this is achieved by separating core functionality from implementation of specific features using an extension mechanism.

Extensions are installed on the host running the VISPA server instance and are typically available for every user who has access to the server.
They are composed of three parts: code and graphics components that extend the client (JavaScript/HTML/CSS), code that handles the communication on the server (Python), and, optionally, code that is transferred to and executed on workspaces (Python).
The purpose of the latter part is to outsource and encapsulate actual extension logic causing the server to take on the role of a mediator between requests and resources.

Included basic extensions are a file browser, a file selector, a code editor, and a terminal.
As they are essential for many common workflows, they are pre-installed on the VISPA server.
Due to the vast development of web techniques over the past years as well as the abundant supply of existing web-based software packages, even more complex and specialized extensions become feasible.
Possible applications range from visualization of static data to highly interactive interfaces, such as graphical steering of modular analyses, which is presented below.

\section{Analysis Designer and Data Browser}\label{sec:pxl}

The extensions presented in this section rely on the \textit{Physics eXtension Library} (PXL), which is a collection of C++ classes addressing analyses in high-energy physics (HEP) and astroparticle physics \cite{pxl}.
It provides basic objects such as the \texttt{pxl::LorentzVector} and the \texttt{pxl::Event} classes, as well as a module system for structuring complex analysis code into reusable building blocks.
Modules can be written in both, C++ and Python.
Objects stored in \textit{pxlio} files can be nested and connected to each other via \textit{ownership} and \textit{relations}.
They also have the ability to store \textit{user records}, i.e., key-value pairs of arbitrary data types.
Therefore, PXL provides a consistent interface to objects and containers, which can be organized in various experiment-specific setups.

Chains of analysis modules can be created graphically using the \textit{Analysis Designer} (fig. \ref{fig:extensions}a).
Users can add, remove and arrange modules using drag and drop gestures.
Options give rise to code flexibility beyond compile time.
Connections between modules control the data flow during execution initiated via the \texttt{pxlrun} command.

Input and output files of these analysis chains can be inspected by the \textit{Data Browser} (fig. \ref{fig:extensions}b). In the case of HEP data, browsing is performed on a per-event basis.
Different views on the event and its contents are presented in dedicated windows.
Currently, Feynman-like graphs, plain content listings and detailed property listings are supported.
However, the code structure allows to add further views such as, e.g., a three-dimensional event display.
Based on the non-restrictive design of PXL objects as outlined above, event content stored in other data formats is convertible to \textit{pxlio} and thus, visualizable by the extension.

\section{Embedding External Functionality: JSROOT}\label{sec:jsroot}

The number of portings of successful desktop software projects to the web increases rapidly in the last years.
A physics-related example is the ROOT framework \cite{root}.
In ROOT, one and two dimensional histograms as well as flat data structures can be visualized with a high degree of interactivity in the \texttt{TBrowser}.
This feature was ported to the web using JavaScript and HTML, and released as JSROOT \cite{jsroot}.
It is even capable of rendering three dimensional objects using \textit{WebGL}.
In order to visualize data, software installation and target files have to be located on the same machine. In certain cases, target files need to be copied to the software location or vice versa.

\begin{figure}[htbp]
\vspace{-6mm}
\begin{subfigure}[b]{0.49\textwidth}
\includegraphics[width=\textwidth]{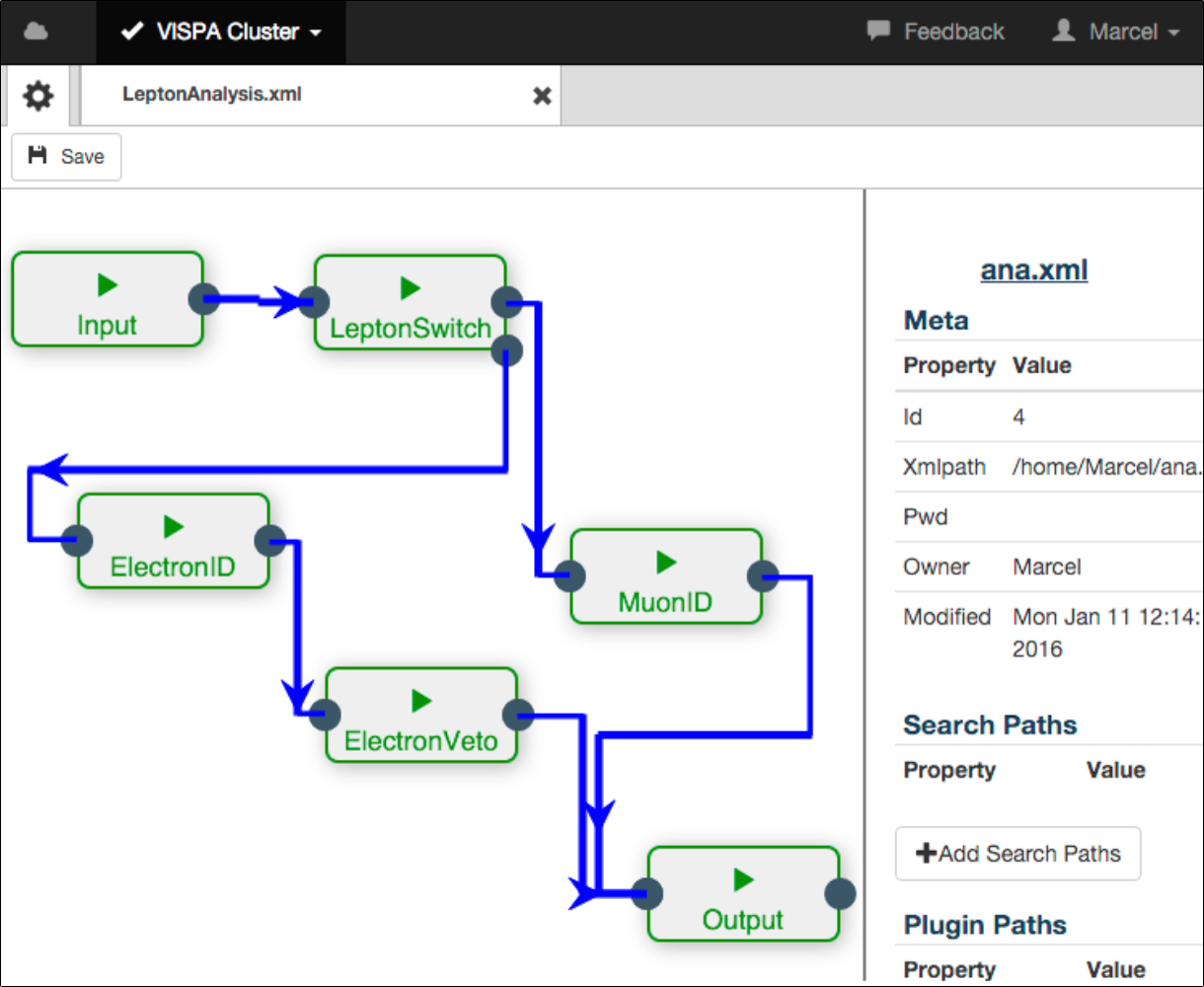}
\subcaption{}
\end{subfigure}
\hfill
\begin{subfigure}[b]{0.49\textwidth}
\includegraphics[width=\textwidth]{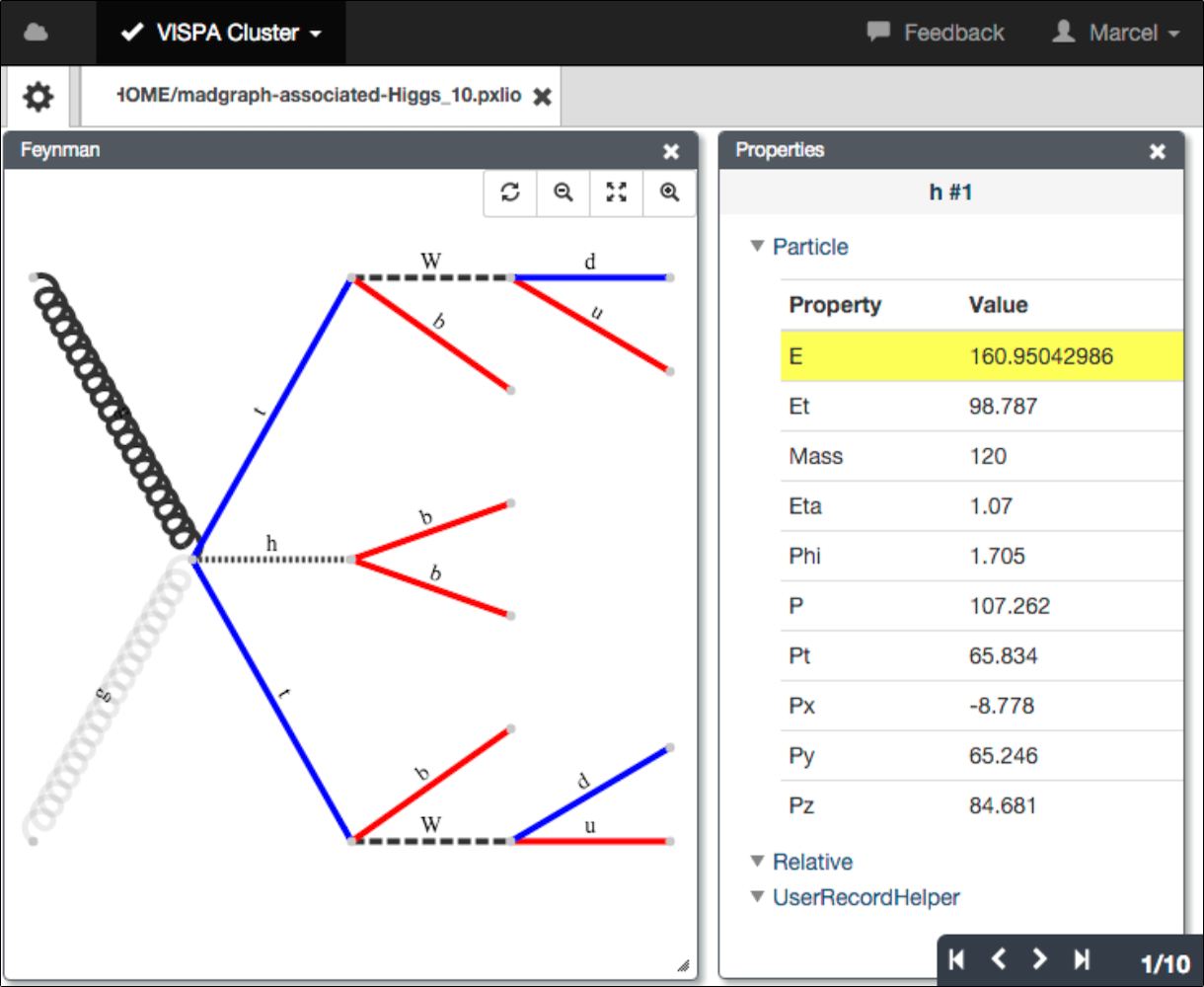}
\subcaption{}
\end{subfigure}\\\\
\begin{subfigure}[b]{\textwidth}
\centering
\includegraphics[width=0.49\textwidth]{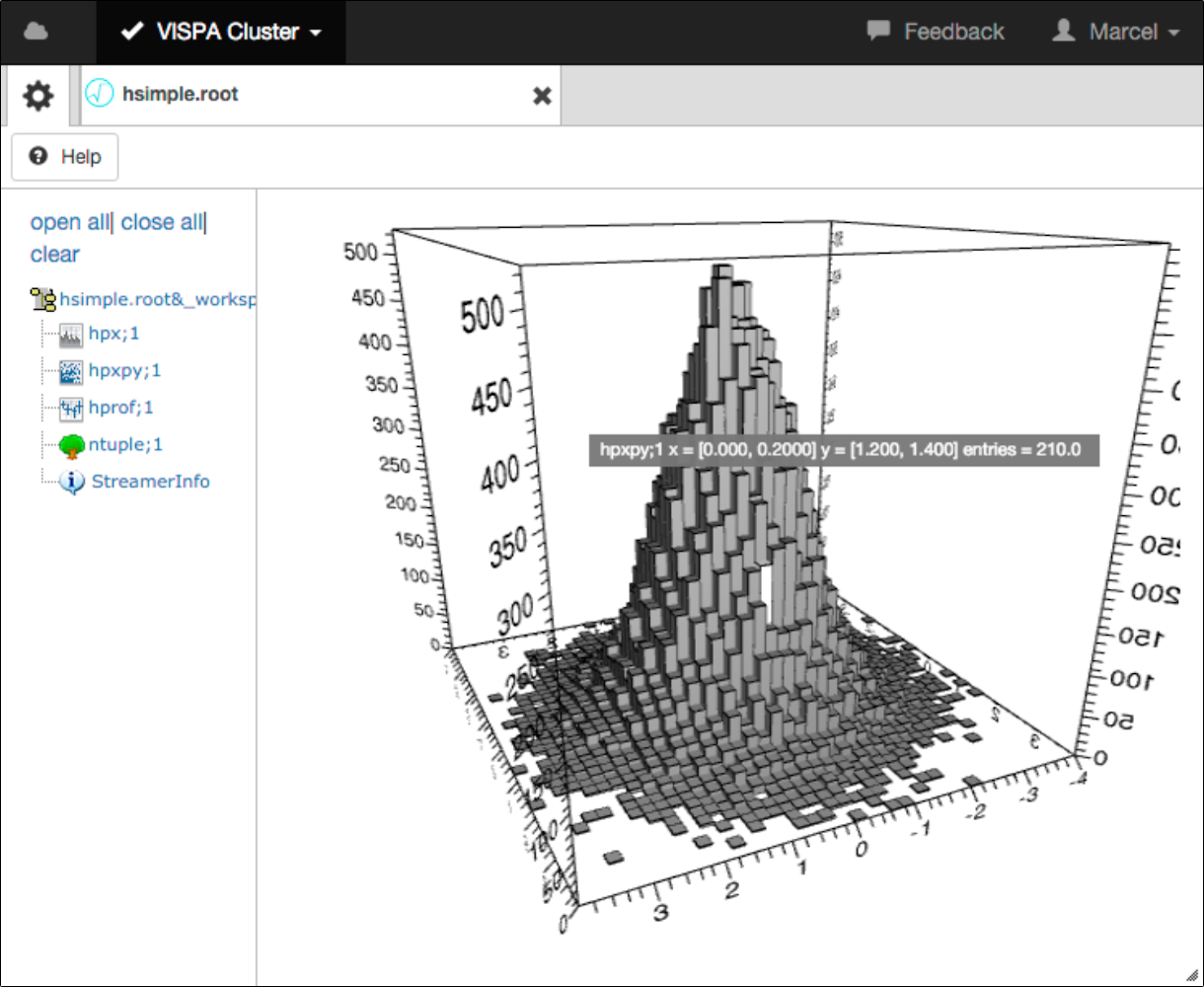}
\subcaption{}
\end{subfigure}
\caption{Analysis-specific extensions of the web-based implementation of VISPA. a) Analysis Designer, b) Data Browser, and c) JSROOT.}
\label{fig:extensions}
\vspace{-4mm}
\end{figure}

As explained in section \ref{sec:software}, VISPA simplifies remote data access using workspaces while requiring only one software installation on a single VISPA server instance.
Technically, JSROOT can be used as a plugin.
We developed a VISPA extension consisting of only a thin code layer that embeds this plugin, which now benefits from exchangeable workspaces (fig. \ref{fig:extensions}d).
It supplies the same features and rich interactions as the standalone JSROOT application.

\section{User Management and Permission System}\label{sec:um}

VISPA provides a server-side user management and permission system, which is configurable via the web-based user interface.
Its purpose is to define groups of users and assign them to customizable sets of permissions in order to control access to resources in a generic fashion.

\textit{Groups} are named collections of users and other groups that are related by similar affiliation and privileges.
\textit{Permissions} describe the privilege to create, access, edit or delete a generic \textit{resource}.
A resource might be a particular file or directory on a workspace, but even more abstract utilizations are possible, such as the configuration of a workspace itself.
VISPA comes with a predefined set of permissions and resources that can be extended or altered to cover arbitrary use cases and applications.
Analogous to groups, permissions can be combined into so-called \textit{roles}, which simplify the assignment of multiple permissions to the same group.
These assignments define the actual user permissions and are bound to specific contexts, called \textit{projects}, handling potential workspace dependence.
As an example, if a resource (e.g. a file or the software to support an extension) is only available on a specific workspace, the permission assignment is only useful in the scope of the associated project.
This approach enables scalable structures that meet requirements ranging from courses using blended learning concepts to large collaborations.

\section{Conclusions}\label{sec:concl}

The VISPA project provides an ecosystem for experiment-specific software.
Its architecture can be considered a ``glue'' between multiple web-based applications and combines them into a single, customizable environment.
Custom extensions have been developed to address several use cases for physics analyses.

The event content of HEP data files can be interactively visualized with the \textit{Data Browser}.
The \textit{Analysis Designer} provides a graphical interface to modular analysis chains based on PXL, which provides a collection of classes which support users performing HEP analyses.
With the example of JSROOT, we showed that the extension mechanism is also capable of embedding existing software.
Executing JSROOT through VISPA enhances flexibility as it benefits from the concept of workspaces.
The new user management provides a flexible, role-based permission system to the platform.
This makes VISPA suitable for situations were strict rules for access are needed, e.g. class rooms were tutors grade students' homework or where access to experiment data is only granted to collaborators.

\section*{Acknowledgments}

We wish to thank the conference organizers for their kind support.
This VISPA project is supported by the Ministerium f\"ur Wissenschaft und Forschung, Nordrhein-Westfalen, the Bundesministerium f\"ur Bildung und Forschung (BMBF), the Helmholz Alliance Physics at the Terascale, and the Deutsche Forschungsgemeinschaft.

\end{document}